\documentclass[10pt]{article}
\usepackage{epsfig,amsmath,graphicx,amssymb,overpic,setspace}
\usepackage{mathrsfs}
\usepackage{epsfig,amsmath,graphicx,amssymb,overpic,url,cite}
\usepackage{epstopdf}
\usepackage{colordvi}
\usepackage{color}
\def\be{\begin{equation}}
\def\ee{\end{equation}}
\def\bee{\begin{eqnarray}}
\def\ene{\end{eqnarray}}
\def\bes{\begin{subequations}}
\def\ees{\end{subequations}}

\newcommand{\PT}{ \mathcal{PT}}

\def\be{\begin{equation}}
\def\ee{\end{equation}}
\def\bee{\begin{eqnarray}}
\def\ene{\end{eqnarray}}
\def\bes{\begin{subequations}}
\def\ees{\end{subequations}}

\begin{document}

\baselineskip=8pt
\renewcommand {\thefootnote}{\dag}
\renewcommand {\thefootnote}{\ddag}
\renewcommand {\thefootnote}{ }

\pagestyle{plain}

\baselineskip=12pt
{\large \bf \noindent Solitonic dynamics and excitations of the nonlinear Schr\"odinger equation with third-order dispersion in non-Hermitian $\PT$-symmetric potentials} \\


\noindent {\bf Yong Chen \& Zhenya Yan}$^\dag$\footnote{$^\dag$ Correspondence and
requests for materials
should be addressed to Z.Y. (zyyan@mmrc.iss.ac.cn)}  \\[0.1in]
{\small Key Laboratory of Mathematics Mechanization, Institute
of Systems Science, AMSS,   Chinese Academy of Sciences, Beijing
100190, China \vspace{0.1in} \\ Dated: Jan. 2016, Sci. Rep.  {\bf 6}, 23478 (2016)}


\vspace{0.15in}

{\baselineskip=12pt


\noindent {\bf Solitons are of the important significant in many fields of nonlinear science such as nonlinear optics, Bose-Einstein condensates, plamas physics, biology, fluid mechanics, and etc.. The stable solitons have been captured not only theoretically and experimentally in both linear and nonlinear Schr\"odinger (NLS) equations in the presence of non-Hermitian potentials since the concept of the parity-time ($\PT$)-symmetry was introduced in 1998. In this paper, we present novel bright solitons of the NLS equation with third-order dispersion in some complex $\PT$-symmetric potentials (e.g., physically relevant $\PT$-symmetric Scarff-II-like and harmonic-Gaussian potentials). We find stable nonlinear modes even if the respective linear $\PT$-symmetric phases are broken. Moreover, we also use the adiabatic changes of the control parameters to excite the initial modes related to  exact solitons to reach stable nonlinear modes. The elastic interactions of two solitons are exhibited in the third-order NLS equation with $\PT$-symmetric potentials. Our results predict the dynamical phenomena of soliton equations in the presence of third-order dispersion and $\PT$-symmetric potentials arising in nonlinear fiber optics and other physically relevant fields.}

\vspace{0.15in}

The representative nonlinear Schr\"odinger (NLS) equation can be used to describe distinguishing wave phenomena arising in many nonlinear physical fields such as nonlinear optics, Bose-Einstein condensates (alias Gross-Pitaevskii equation), the deep ocean, DNA, plasmas physics, and even financial market, etc.~\cite{nls,nls2,nls3,nls3a,nls3b,ss3,bec,yanctp10,yanpla11}. The dynamics of fundamental bright and dark solitons (also vortices and light bullets in higher-dimensional cases) of the NLS model and self-similar modes in its generalized forms have been addressed (see Refs.~\cite{s1,s2,s3,s3a,s4,s5,s6,s7} and references therein). Recently,  inspired by the {\em non-Hermitian} $\PT$-symmetric potentials first suggested by Bender and Boettcher~\cite{bender1,bender2} in the classical Hamiltonian operators,  Musslimani {\it et al.}~\cite{ziad} first introduced the complex $\PT$-symmetric potentials in the NLS model such that some novel phenomena with stable modes were found. After that, a variety of distinguishing $\PT$-symmetric or non-$\PT$-symmetric potentials were introduced to the continuous or discrete NLS equations to explore dynamical behaviors of $\PT$-symmetric nonlinear modes (see, e.g., Refs.~\cite{pt0,pt1,pt2,pt3,pt4,pt5,pt6,pt7, pt72,pt73, pt81,pt82,pt83,pt84,pt8,pt9,pt10,pt10a,pt11,pt12,pt13,pt14} and references therein). Meanwhile, more and more physical experiments have also been designed to observe new wave phenomena in the sense of non-Hermitian $\PT$-symmetric potentials~\cite{exp1,exp2,exp3,exp4,exp6,exp7}. Here the parity ${\cal P}$ and temporal ${\cal T}$ operators are defined as~\cite{bender2}: ${\cal P}:\, {\cal P} \,p\,{\cal P}=-p , \, {\cal P}\, x\,{\cal P}= - x$ and ${\cal T}:\, {\cal T}\, p\,{\cal T}=-p, \, {\cal T}\, x\,{\cal T}= x,\, \, {\cal T}\, i\,{\cal T}=-i$.
The one-dimensional complex potential $U(x)$ is $\PT$-symmetric provided that the sufficient (not necessary) conditions $U_{R}(x)=U_{R}(-x)$ and $U_{I}(-x)=-U_{I}(x)$ hold~\cite{bender2}, where $U(x)$ is also called the refractive-index in optical fibre.

In the study of ultra-short (e.g., $100$ fs ~\cite{nls}) optical pulse propagation, the higher-order dispersive and nonlinear effects become significant such as third-order dispersion (TOD), self-steepening (SS), and the self-frequency shift (SFS)
arising from the stimulated Raman scattering. The third-order NLS equation was introduced from the Maxwell equation~\cite{ss,ss2}. The generalized inhomogeneous third-order NLS equation with modulating coefficients in the complex gain-or-loss term has been verified to admit optical rogue waves~\cite{yanjop}.  Recently, the NLS equation with only third-order dispersion was used to numerically confirm the experimental  observation of the spectral signature of the collision between a soliton and the dispersive wave~\cite{3nls3}. To our best knowledge, the  $\PT$-symmetric linear and nonlinear modes in the third-order NLS equation were not studied before. Our aim in this paper is to investigate the linear and nonlinear modes of the third-order NLS equation in the presence of physically interesting $\PT$-symmetric potentials, e.g., Scarff-II-like potential and harmonic-Gaussian potential. We find that some parameters can modulate the stable nonlinear modes even if the linear $\PT$-symmetric phases are broken. Moreover, we also understand that the adiabatic changes of control parameters can be used to excite the initial modes subject to exact bright solitons to generate
stable nonlinear modes.

The rest of this report is arranged as follows. In Section of Results,  we introduce the NLS equation with third-order dispersion in the presence of complex $\PT$-symmetric potentials. We consider the nonlinear modes and their stability in the $\PT$-symmetric Scarff-II-like and harmonic-Gaussian potentials. The problems of nonlinear modes excitations is also investigated, which can excite initial nonlinear modes to reach stable modes. Moreover, we also give some methods used in this paper. Finally, some conclusions and discussions are presented.
\vspace{0.1in}

\noindent{\large\bf Results}

\noindent{\bf Nonlinear wave model with $\PT$-symmetric potentials.} We focus on the generalized form of the third-order NLS equation~\cite{3nls1, 3nls2,3nls3} in non-Hermitian potentials, that is, the NLS equation with third-order dispersion (TOD) and complex $\PT$-symmetry potentials
\bee\label{nls}
 i\frac{\partial\psi}{\partial z}=-\frac{1}{2}\frac{\partial^2\psi}{\partial x^2}-i\frac{\beta}{6}\frac{\partial^3\psi}{\partial x^3}+[V(x)+iW(x)]\psi-g|\psi|^2\psi
\ene
where Raman effect, nonlinear dispersion terms (e.g., self-steepening term and self-frequency shift effect), and higher-order dispersion terms are neglected\cite{raman,ss,ss2,ss4}, $\psi\equiv \psi(x, z)$ is a complex wave function of $x,z$, $z$ denotes the propagation distance, the real parameter $\beta$ stands for the coefficient of TOD, the $\PT$-symmetric potential requires that $V(x)=V(-x)$ and $W(x)=-W(-x)$ describing the real-valued external potential and gain-and-loss distribution, respectively, and $g>0$\, (or $<0)$ is real-valued inhomogeneous self-focusing (or defocusing) nonlinearity. The power of Eq.~(\ref{nls}) is given by $P(z)=\int_{-\infty}^{+\infty}|\psi(x,z)|^2dx$ and  one can
readily know that $P_z=2\int_{-\infty}^{+\infty}W(x)|\psi(x,z)|^2dx$.  Eq.~(\ref{nls}) becomes the usual higher-order NLS equation in the absence of the gain-and-loss distribution~\cite{3nls1}. Eq.~(\ref{nls}) with $\beta=0$ becomes the $\PT$-symmetric nonlinear model, which has been studied~\cite{pt0,pt1,pt2,pt3,pt4,pt5,pt6,pt7, pt72,pt73, pt8,pt81,pt82,pt83,pt84,pt9,pt10,pt10a,pt11,pt12}. In the following we consider the case in the presence of TOD term ($\beta\not=0$) and gain-and-loss distribution. Here our following results are  also suitable for the case $x\to t$ in Eq.~(\ref{nls}).

\vspace{0.1in}
\noindent{\bf Linear spectrum problem with $\PT$-symmetric potential.} We start to study the physically interesting potential in Eq.~(\ref{nls}) as the $\PT$-symmetric Scarff-II-like potential
\bee \label{gps}
 V(x)=V_0\,{\rm sech}^2 x, \quad  W(x)=\beta\, {\rm sech}^2x\tanh x,
\ene
where the real constant $V_0$ and TOD parameter $\beta$ can be used to modulate the amplitudes of the reflectionless potential $V(x)$~\cite{rp} and gain-and-loss distribution $W(x)$, respectively. Moreover, $V(x)$ and $W(x)$ are both bounded (i.e., $0<|V(x)|\leq |V_0|,\, |W(x)|\leq 2\sqrt{3}|\beta|/9$) with $W(x)=\beta V_0^{-2}V(x)\sqrt{V_0^2-V^2(x)}$ and $V(x), \, W(x) \to 0$ as $|x|\to \infty$ (see Fig.~\ref{fig1}a). It is easy to see that the gain-and-loss distribution is always balanced since $\int_{-\infty}^{+\infty}W(x)dx=0$ and  has only the limit effect on linear and nonlinear modes since $W(x)\sim 0$ as $|x|>M>0$.   The sole difference between the potential (\ref{gps}) and the usual Scraff-II potential~\cite{sc} is that the gain-and-loss distribution in Eq.~(\ref{gps}) will more quickly approaches to zero than one (i.e., $\beta {\rm sech}\,x\tanh x $) in Scarff-II potential for the same amplitudes.

We firstly consider the linear spectrum problem (i.e., Eq.~(\ref{nls}) with  $g=0$) in the $\PT$-symmetric Scarff-II-like potential (\ref{gps}) and use the stationary solution transformation $\psi(x,z)=\Phi(x)e^{-i\lambda z}$ to yield
\bee \label{ls}
 L\Phi(x)=\lambda\Phi(x),\qquad L=-\frac12\frac{d^2}{dx^2}-i\frac{\beta}{6} \frac{d^3}{dx^3}+V(x)+i W(x)
\ene
where $\lambda$ and $\Phi(x)$ are the corresponding  eigenvalue and eigenfunction, respectively, and $\lim_{|x|\to \infty}\Phi(x)=0$. Since the discrete spectrum
of a complex $\PT$-symmetric potential is either real or appears in complex conjugated pairs, thus we may find some proper parameters $V_0,\, \beta$ for which the complex $\PT$-symmetric potential keep unbroken.

Here we consider $V_0<0$ such that the shape of potential $V(x)$ seems to be $V$-shaped with zero boundary conditions (see Fig.~\ref{fig1}a). We numerically study the discrete spectra of the operator $L$ (see Methods).  Fig.~\ref{fig1}b exhibits the regions of broken and unbroken $\PT$-symmetric phases on the $(V_0,\, \beta)$ space. Two almost parallel straight lines ($\beta\approx\pm 0.12$)  separate the limited space $\{(V_0, \beta) | -0.02\leq V_0\leq -3,\, |\beta|\leq 0.5\}$. The regions of broken and unbroken $\PT$-symmetric phases are outside and inside between two lines, respectively. For the given TOD parameter $\beta=0.1$ and varying $V_0$, the spontaneous symmetry breaking does not occur from the two lowest states since the maximum absolute value of imaginary parts of $\lambda$ is less than $6\times 10^{-14}$ and they can be regarded as zero (see Figs.~\ref{fig1}(c,d)).
However, for the given $V_0=-2$ and varying $\beta$, the spontaneous symmetry breaking occurs from two lowest states starting from some $\beta=0.12$ (see Figs.~\ref{fig1}(e,f)).

\begin{figure}[!t]
	\begin{center}
		\hspace{-0.05in}{\scalebox{0.55}[0.55]{\includegraphics{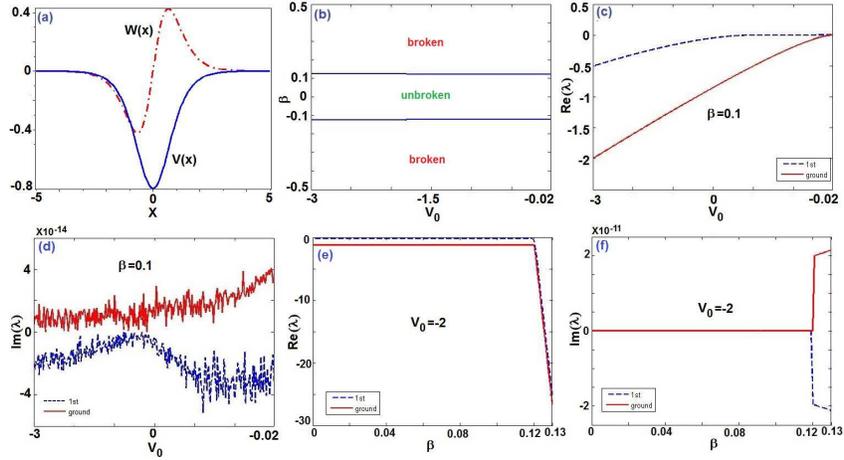}}}
				\end{center}
	\vspace{-0.15in}
	\caption{{\bf  Linear spectrum problem.} (a) $\PT$-symmetric potential (\ref{gps}) with $V_0=-0.8,\, \beta=1$, (b) The phase transitions for the linear operator $L$ (\ref{ls}) with  $\PT$-symmetric Scarff-II-like potentials (\ref{gps}). The domain of unbroken (broken) $\PT$-symmetric phase is inside (outside) the domain in $(V_0, \beta)$-space. (c) Real and (d) imaginary parts of the eigenvalues $\lambda$ [see Eq.~(\ref{ls})] as functions of $V_0$ for the $\PT$-symmetric potential (\ref{gps}) at $\beta=0.1$, in which the imaginary parts are almost zero.  (e) Real and (f) imaginary parts of the eigenvalues $\lambda$ [see Eq.~(\ref{ls})] as functions of $\beta$ for the $\PT$-symmetric potential (\ref{gps}) at $V_0=-2$.} \label{fig1}
\end{figure}

\vspace{0.1in}
\noindent{\bf Nonlinear localized modes and stability.} For the given $\PT$-symmetric Scarff-II-like potential (\ref{gps}), based on some transformations, we can find the unified analytical bright solitons of Eq.~(\ref{nls}) for both the self-focusing and defocusing cases (see Methods)
\bee\label{solu}
 \psi(x,z)=\sqrt{(V_0-\kappa\beta+1)/g}\,{\rm sech} (x) e^{i\kappa x-i\mu z},
\ene
where the phase wavenumber is defined by TOD coefficient $\kappa=(1+\nu\sqrt{1+\beta^2/3})/\beta$ with  $\nu=\pm 1$, the potential is $\mu=(3\kappa^2+3\beta \kappa-\beta \kappa^3-3)/6$, and
 the existent condition $g(V_0-\kappa\beta+1)>0$ is required. For the signs of parameter $\nu$ and nonlinearity $g$, we find the following four cases for the existent conditions of bright solitons (\ref{solu}) (let $\alpha=\sqrt{1+\beta^2/3}$):
(i) $\nu=-g=-1$ and $V_0>-\alpha$ (i.e., the right-side domain of the hyperbola of one sheet $V_0=-\alpha$ on $(V_0,\, \beta)$-space); (ii) $\nu=g=-1$ and $V_0<-\alpha$ (i.e., the left-side domain of the hyperbola of one sheet $V_0=-\alpha$ on $(V_0,\, \beta)$-space); (iii) $\nu=g=1$ and $V_0>\alpha$ (i.e., the right-side domain of the hyperbola of one sheet $V_0=\alpha$ on $(V_0,\, \beta)$-space); (iv) $\nu=-g=1$ and $V_0<\alpha$ (i.e., the left-side domain of the hyperbola of one sheet $V_0=\alpha$ on $(V_0,\, \beta)$-space).

In the following we numerically~\cite{yang} study the robustness (linear stability) of nonlinear localized  modes (\ref{solu}) for both self-focusing and defocusing cases ($g=\pm 1$) via the direct propagation of the initially stationary state (\ref{solu}) with a noise perturbation of order  about $2\%$ in Fig.~\ref{fig2} (see Methods). Fig.~\ref{fig2}(a1)
exhibits the stable and unstable (approximate) regions for Case (i) $\nu=-g=-1$ and different parameters $V_0$ and $\beta$. For $V_0=-0.8,\, \beta=0.1$ belonging to the domain of the unbroken linear $\PT$-symmetric phase [cf. Eq.~(\ref{ls}) and Fig.~\ref{fig1}b], the stable nonlinear mode is generated (see Fig.~\ref{fig2}(a2)). For the fixed $V_0=-0.8$, if we change $\beta=1.1$ corresponding to the domain of the broken linear $\PT$-symmetric phase, then a stable nonlinear mode is found too, that is, the focusing nonlinear term can modulate the unstable linear modes (broken $\PT$-symmetric phase) to stable nonlinear modes (see Fig.~\ref{fig2}(a3)). But if $\beta$ increases a little bit (e.g., $\beta=1.5$), then the nonlinear mode becomes unstable (see Fig.~\ref{fig2}(a4)). In particular, for $V_0=-1.1,\, \beta=0.7$, which does not satisfy the required existent condition of solution $V_0>-\alpha$, that is, the expression (\ref{solu}) for this case, $\psi_0(x,z)=i\sqrt{1.1-\sqrt{3.49/3}}\,{\rm sech}(x)e^{i\kappa x-i\mu z}$, is not an analytical solution of Eq.~(\ref{nls}), but we still use it as the initial solution with a noise perturbation of order $2\%$ to make numerical simulations such that we find the initial mode $\phi_{0}(x,0)$ can be excited to a stable and weakly oscillatory (breather-like behavior) situation (see Fig.~\ref{fig2}(a5)). For Case (ii) $\nu=g=-1$, we also have the similar results (see Figs.~\ref{fig2}(b3)-(b5)). For the last two Cases (iii) and (iv), we fix $V_0>0$, in which the potential is similar to a Gaussian-like profile and the linear problem (\ref{ls}) has no discrete spectra, but we still find the stable nonlinear modes (see Figs.~\ref{fig2}(c3), (c4), (c5), (d3), (d5)) and unstable (see Figs.~\ref{fig2}(c2), (d2), (d4)) nonlinear modes.

 For the above-obtained nonlinear modes (\ref{solu}), we have the corresponding transverse power-flow or Poynting vector given by $S=i/2(\psi\psi_x^{*}-\psi^{*}\psi_x)=\kappa g^{-1}(V_0-\kappa\beta+1)\,{\rm sech}^2x$ with $(V_0-\kappa\beta+1)/g>0$ and $\kappa=(1+\nu\sqrt{1+\beta^2/3})/\beta$. We here consider the only case $\beta>0$ (the case $\beta<0$ can also be considered). For $\nu=1$\, (or $-1)$, we have $S>0$ (or $<0)$, which implies that the pamaeter $\nu$ change the directions of power flows from gain to loss regions.
 The power of the solutions (\ref{solu}) is $P(z)=\int_{-\infty}^{+\infty}|\psi(x,z)|^2dx=2(V_0-\kappa\beta+1)/g$, which is conserved.

\begin{figure}[!t]
	\begin{center}
		\hspace{-0.05in}{\scalebox{0.7}[0.65]{\includegraphics{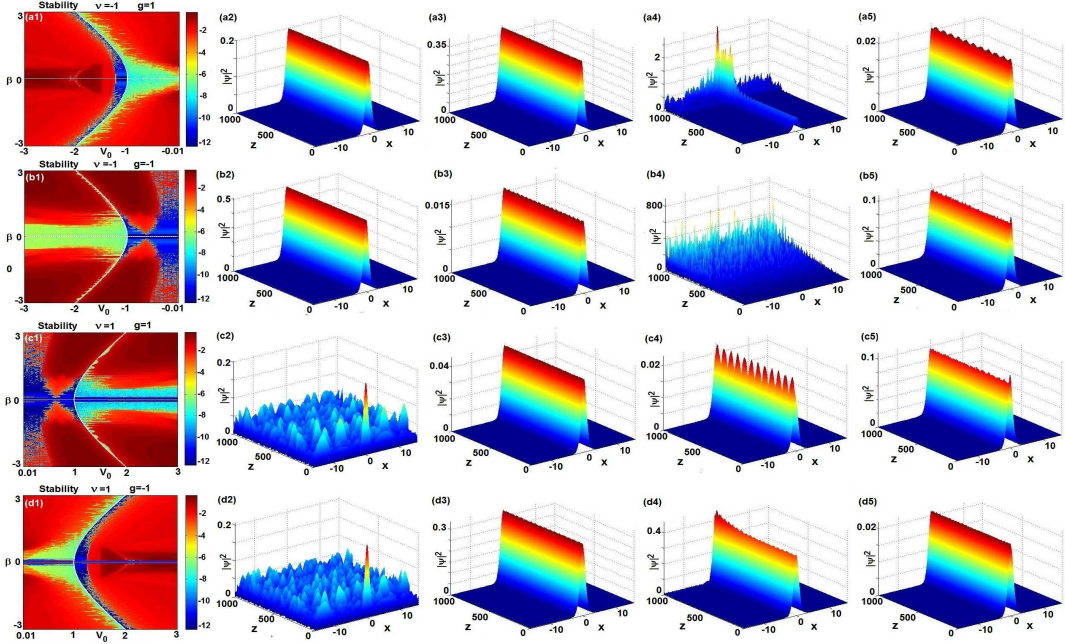}}}
				\end{center}
	\vspace{-0.15in}
	\caption{{\bf Stability of nonlinear modes (\ref{solu}).}
(a1)-(a5) $\nu=-g=-1$, (a1) stable and unstable regions [the maximal absolute value of imaginary parts of the linearized eigenvalue $\delta$ on $(V_0, \beta)$ space (common logarithmic scale), similarly hereinafter], (a2) $V_0=-0.8,\, \beta=0.1$ (stable), (a3) $V_0=-0.8,\, \beta=1.1$ (stable), (a4) $V_0=-0.8,\, \beta=1.5$ (unstable), (a5) $V_0=-1.1,\, \beta=0.7$ (periodically varying);
(b1)-(b5) $\nu=g=-1$, (b2) $V_0=-1.5,\, \beta=0.1$ (stable), (b3) $V_0=-1.5,\, \beta=1.9$ (stable), (b4) $V_0=-1.5,\, \beta=2$ (unstable), (b5) $V_0=-0.9,\, \beta=0.2$ (stable);
(c1)-(c5) $\nu=g=1$,  (c2) $V_0=1.2,\, \beta=0.1$ (unstable), (c3) $V_0=1.2,\, \beta=1$ (stable), (c4) $V_0=1.2,\, \beta=1.2$ (periodically varying), (c5) $V_0=0.9,\, \beta=0.2$ (stable);
(d1)-(d5) $\nu=-g=1$, (d2) $V_0=0.8,\, \beta=0.1$ (unstable), (d3) $V_0=0.8,\, \beta=1$ (stable), (d4) $V_0=0.8,\, \beta=1.1$ (unstable), (d5) $V_0=1.1,\, \beta=0.7$ (stable).   } \label{fig2}
\end{figure}

\begin{figure}[!t]
	\begin{center}
		\hspace{-0.05in}{\scalebox{0.75}[0.75]{\includegraphics{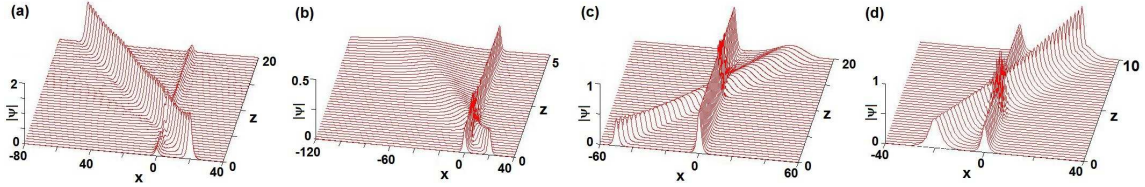}}}
				\end{center}
	\vspace{-0.15in}
	\caption{{\bf The interactions of bright solitons (\ref{solu}) of Eq.~(\ref{nls}).} (a) $V_0=1.1,\, \beta=\nu=-g=1$, (b) $V_0=1.2,\, g=\beta=\nu=1$, (c)
$g=\nu=-1,\,V_0=-1.5,\, \beta=0.1$, (d) $g=-\nu=1,\,V_0=-0.8,\, \beta=0.1$.} \label{coll}
\end{figure}

\begin{figure}[!t]
	\begin{center}
		\vspace{0.05in}{\scalebox{0.56}[0.56]{\includegraphics{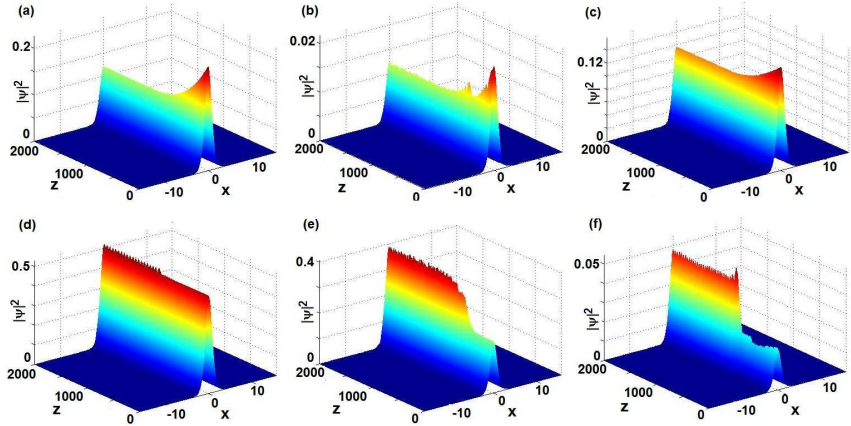}}}
				\end{center}
	\vspace{-0.15in}
	\caption{{\bf Exciting stable nonlinear localized modes [cf. Eq.~(\ref{nls2})].} (a) $\nu=-g=1,\, V_0=0.8,\, \beta_1=0.4,\, \beta_2=0.1$,
  (b) $\nu=-g=1,\, V_0=0.8,\, \beta_1=0.7,\, \beta_2=0.3$, (c) $\nu=g=1,\, V_0=1.2,\, \beta_1=0.5,\, \beta_2=0.1$,
  (d) $\nu=g=-1,\, V_0=-1.5,\, \beta_1=0.6,\, \beta_2=0.1$, (e) $\nu=-g=-1,\, V_0=-0.8,\, \beta_1=1,\, \beta_2=0.1$,
  (f)  $\nu=-g=-1,\, V_0=-1.1,\, \beta_1=0.7,\, \beta_2=0.1$.
   } \label{excite}
\end{figure}

\vspace{0.1in}
\noindent{\bf Interactions of bright solitons.} We here study the interactions of two bright solitons in the $\PT$-symmetric potential. For the defocusing case $g=-1$, if we choose $V_0=1.1,\, \beta=\nu=1$ and consider the initial condition $\psi_1(x,0)=\phi(x, 0)+\sqrt{1.2+2/\sqrt{3}}\,{\rm sech}(x-20)e^{4ix}$ with $\phi(x, 0)$ given by Eq.~(\ref{solu}) such that the elastic interaction is generated (see Fig.~\ref{coll}a). If we choose $V_0=-1.5,\, \beta=0.1,\, \nu=-1$ and consider the initial condition $\psi_2(x,0)=\phi(x, 0)+ 0.5\,{\rm sech}(0.5x+10)e^{7ix}$
with $\phi(x, 0)$ given by Eq.~(\ref{solu}) such that the elastic interaction is generated (see Fig.~\ref{coll}c). For the self-focusing case $g=1$,  if we choose $V_0=1.2,\, \beta=\nu=1$ and consider the initial condition $\psi_3(x,0)=\phi(x, 0)+\sqrt{2/\sqrt{3}-1.1}\,{\rm sech}(x-20)e^{7ix}$ with $\phi(x, 0)$ given by Eq.~(\ref{solu}) such that the elastic interaction is generated (see Fig.~\ref{coll}b).  If we choose $V_0=-0.8,\, \beta=0.1,\, \nu=-1$ and consider the initial condition $\psi_4(x,0)=\phi(x, 0)+0.5\,{\rm sech}(0.5x+10)e^{7ix}$ with $\phi(x, 0)$ given by Eq.~(\ref{solu}) such that the elastic interaction is generated (see Fig.~\ref{coll}d).

\vspace{0.1in}
\noindent{\bf Exciting stable nonlinear localized modes in equation (\ref{nls}).} Nowadays we turn to the excitation of nonlinear modes by means
of a slow change of the control TOD parameter $\beta\to \beta(z)$ in Eq.~(\ref{nls}) which is regarded as a function of propagation distance $z$, that is, we
focus on simultaneous adiabatic switching on the TOD and gain-and-loss distribution, modeled by
\bee\label{nls2}
 i\frac{\partial\psi}{\partial z}=-\frac{1}{2}\frac{\partial^2\psi}{\partial x^2}-i\frac{\beta(z)}{6}\frac{\partial^3\psi}{\partial x^3}+[V(x)+iW(x)]\psi-g|\psi|^2\psi
\ene
where $V(x), W(x)$ are given by Eq.~(\ref{gps}) with $\beta\to \beta(z)$, and $\beta(z)$ is chosen as
\bee \label{beta}
\beta(z)=\begin{cases} \beta_2\sin(\pi z/2000)+\beta_1, \quad  0\leq z <1000, \\
          \beta_1+\beta_2, \qquad\qquad  \qquad\quad                        1000\leq z
          \end{cases}
\ene
with $\beta_{1,2}$ being real constants. It is easy to see that the solutions (\ref{solu}) with $\beta\to \beta(z)$ do not satisfy Eq.~(\ref{nls2}), but for $z=0$ and $z\geq 1000$ solutions (\ref{solu}) indeed satisfy Eq.~(\ref{nls2}).

Fig.~\ref{excite} displays the wave propagation of the solutions $\psi(x,z)$ of Eq.~(\ref{nls2}) subject to the initial condition
given by Eq.~(\ref{solu}) with $\beta\to \beta(z)$ given by Eq.~(\ref{beta}). For $\nu=1$ and different parameters $g,\, V_0,\, \beta_{1,2}$, Figs.~\ref{excite}(a,b,c) exhibit the stable modes in which the initial states
$|\psi(x,0)|^2$ given by Eq.~(\ref{solu}) with $z=0,\, \beta=\beta_1$ are all of the higher amplitudes and then the amplitudes decrease slowly as $z$ increases such that they reach the alternative stable sates beginning from about $z=1000$. For $\nu=-1$ and different parameters $g,\, V_0,\, \beta_{1,2}$, Figs.~\ref{excite}(e,f) also exhibit the stable modes in which the initial states are all of the lower amplitudes and then the amplitudes grow step and step as $z$ increases such that they reach the stable and weakly oscillatory (breather-like behavior) situations beginning from about $z=1000$, but Fig.~\ref{excite}d shows that the stable mode keeps from $z=0$ to $z=1100$ and then the wave slowly  increases a little bit to reach another stable and weakly oscillatory (breather-like behavior) feature. In particular, in Fig.~\ref{excite}b (or f), we can excite the initial states subject  to inexact solitons (\ref{solu}) of Eq.~(\ref{nls}) for $\nu=-g=1,\, V_0=0.8,\, \beta=0.7$ (or $\nu=-g=-1,\, V_0=-1.1,\, \beta=0.7$) to the stable states subject to exact solitons (\ref{solu}) of Eq.~(\ref{nls}) for $\nu=-g=1,\, V_0=0.8,\, \beta=1$ (or $\nu=-g=-1,\, V_0=-1.1,\, \beta=0.8$ of Eq.~(\ref{nls})).

\begin{figure}[!t]
	\begin{center}
		\hspace{-0.05in}{\scalebox{0.6}[0.6]{\includegraphics{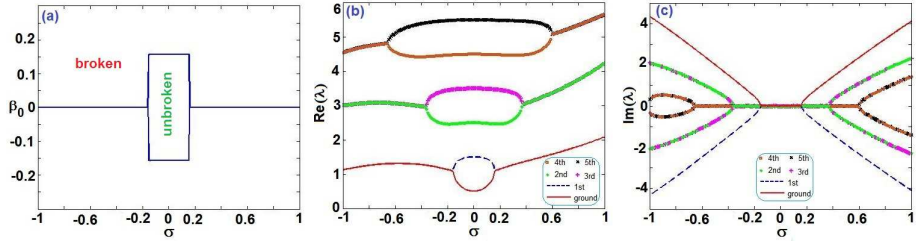}}}
				\end{center}
	\vspace{-0.15in}
	\caption{ {\bf Linear spectrum problem.} (a) The phase transitions for the linear operator $L$ (\ref{ls}) with $\PT$-symmetric harmonic-Gaussian potential (\ref{hg}). The unbroken (broken) $\PT$-symmetric phase is in the domain inside (outside) the almost rectangle domain on $(\sigma, \beta_0)$-space. (b) Real and (c) imaginary parts of the eigenvalues $\lambda$ [see Eq.~(\ref{ls})] as functions of $\sigma$ for the $\PT$-symmetric potential (\ref{hg}) at $\beta_0=0.1$.} \label{fig1-spectra-gau}
\end{figure}

\vspace{0.1in}
\noindent{\bf Nonlinear model with the spatially varying TOD.} We here consider nonlinear modes of the generalized form of Eq.~(\ref{nls}) with $x$-spatially varying TOD coefficient $\beta\to \beta(x)$, that is,
 \bee\label{nlsg}
 i\frac{\partial\psi}{\partial z}=-\frac{1}{2}\frac{\partial^2\psi}{\partial x^2}-i\frac{\beta(x)}{6}\frac{\partial^3\psi}{\partial x^3}+[V(x)+iW(x)]\psi-g|\psi|^2\psi
\ene
We here are interested in the TOD coefficient $\beta(x)$ of a Gaussian function
\bee \label{gau}
 \beta(x)=\beta_0 e^{-x^2}
 \ene
  with $\beta_0\not=0$ being a real amplitude of the Gaussian profile and another complex $\PT$-symmetric harmonic-Gaussian potential
\bee \label{hg}
 V(x)=\frac12 x^2+\beta\gamma(36\gamma^2-7x^2+4), \qquad  W(x)=\frac{1}{6}(3x-x^3)\beta+9x\gamma(4\beta \gamma-1),
\ene
in which $\beta=\beta(x)$ is given by Eq.~(\ref{gau}) and we have introduced the function $\gamma=\sigma e^{-x^2/2}$ with $\sigma\not=0$ being a real constant. In fact, we can also consider the general case $\beta(x)$. We know that the potential $V(x)$ approaches to the harmonic potential $x^2/2$ (which is related to the usual physical experiments) and $W(x)\to 0$ as $|x|\to \infty$ (see Figs.~\ref{stability-gau}(b,d)).

For the given $\PT$-symmetric harmonic-Gaussian potential (\ref{hg}), we studied the discrete spectra of linear problem (\ref{ls}) such that we give the separated regions for the broken and unbroken $\PT$-symmetric phase (see Fig.~\ref{fig1-spectra-gau}(a)). When $|\sigma|$ is greater than about $0.16$, we only find the discrete spectra for $\beta_0$ very approaching to zero (here we consider $ |\beta_0|\leq 0.3$). For the given amplitude $\beta_0=0.1$ of  TOD coefficient, we give the first six lowest eigenvalues (see Figs.~\ref{fig1-spectra-gau}(b,c)), where the spontaneous symmetry breaking occurs from the two lowest states.

We find the exactly analytical solutions of Eq.~(\ref{nlsg}) with the Gaussian TOD coefficient in the $\PT$-symmetric harmonic-Gaussian potential (\ref{hg}) in the form
\bee\label{solu2}
 \psi(x, z)=3\sigma\sqrt{2/g}\,e^{-x^2/2}e^{-iz/2+3i\sigma\sqrt{2\pi}\,{\rm erf}(x/\sqrt{2})},
\ene
where $g>0$ and ${\rm erf}(\cdot)$ is an error function. In the following we take $g=1$ without loss of generality.

In what follows we numerically investigate the robustness (linear stability) of nonlinear localized  modes (\ref{solu2}) for self-focusing case ($g=1$) via the direct propagation of the initially stationary state (\ref{solu2}) with a noise perturbation of order  about $2\%$ in Fig.~\ref{stability-gau} (see Methods).
The linear stability of the solutions (\ref{solu2}) is displayed in Fig.~\ref{stability-gau}a) such that we find the solutions (\ref{solu2}) are possibly stable nearby $\sigma=0$. For $\sigma=\beta_0=0.1$, in which the potential becomes almost a harmonic potential $x^2/2$ (see Fig.~\ref{stability-gau}b), we find the stable nonlinear mode (see Fig.~\ref{stability-gau}c). But  for $\sigma=0.2,\, \beta_0=0.35$, in which the potential is a double-well potential (see Fig.~\ref{stability-gau}d), we also obtain the stable nonlinear mode (see Fig.~\ref{stability-gau}e).

 For the above-obtained nonlinear modes (\ref{solu2}), we have the corresponding transverse power-flow or Poynting vector given by $S=i/2(\psi\psi_x^{*}-\psi^{*}\psi_x)=108\sigma^3g^{-1}\,e^{-3x^2/2}$, which implies that ${\rm sgn}(S)={\rm sgn}(\sigma)$. Since the gain-and-loss distribution $W(x)$ given by Eq.~(\ref{hg}) depends on TOD parameter $\beta_0$ and $\sigma$, which generate that there are  more one  intervals for gain (or loss) distribution, thus though the power flows along the positive (negative) direction for the $x$ axis as $\sigma>0$ ($<0$), it is of the complicated structures from the gain-and-loss view.

Moreover, we also study the interactions of two bright solitons in the $\PT$-symmetric potential. For the focusing case $g=1$, if we choose
$\sigma=-0.1,\, \beta_0=0.1$ and consider the initial condition $\psi_1(x,0)=\phi(x, 0)-0.3\sqrt{2}e^{-(x-10)^2/2}$ with $\phi(x, 0)$ given by Eq.~(\ref{solu2}) such that the elastic interaction is generated (see Fig.~\ref{coll-gau}(a)). If we choose $\sigma=0.2,\, \beta_0=0.1$ and consider the initial condition $\psi_2(x,0)=\phi(x, 0)+0.3\sqrt{2}e^{-(x-10)^2/2}$
with $\phi(x, 0)$ given by Eq.~(\ref{solu2}) such that the elastic interaction is generated too (see Fig.~\ref{coll-gau}(b)).

Nowadays we turn to the excitation of nonlinear modes by means
of a slow change of the control TOD parameter $\beta(x)\to \beta(x,z)$ in Eq.~(\ref{nls2}) whose amplitude $\beta_0$ is regarded as a function of propagation distance $z$, that is, we focus on simultaneous adiabatic switching on the TOD, potential, and gain-and-loss distribution.

For the given $\sigma=0.1$, we consider the varying amplitude $\beta_0\to \beta_0(z)$ in Eq.~(\ref{gau}) in the form
\bee\label{beta0}
\beta_0(z)=\begin{cases} 0.9\sin(\pi z/2000)+0.1, \quad  0\leq z <1000, \\
           1, \qquad\qquad  \qquad\qquad \qquad                       1000\leq z
          \end{cases}
\ene
which makes the TOD coefficient given by Eq.~(\ref{gau}), potential $V(x)$ and gain-or-loss distribution $W(x)$ given by Eq.~(\ref{hg}) change, but it does not change the expression of solutions (\ref{solu2}). Now we study the wave evolution of the solution $\psi(x,z)$ satisfied by Eq.~(\ref{nls2}) with Eqs.~(\ref{gau}), (\ref{hg}), and (\ref{beta0}) subject to the initial condition given by Eq.~(\ref{solu2}). As a consequence, we find the stable nonlinear modes by using the excitation (\ref{beta0}), that is, we can steadily excite one stable mode (Fig.~\ref{stability-gau}(c)) to reach another stable one (Fig.~\ref{stability-gau}(e)), which is exhibited in Fig.~\ref{excite-gau}(a).

Now we fix the amplitude of TOD, $\beta_0=0.1$, and consider the effect of varying amplitude $\sigma\to\sigma(z)$ on nonlinear modes:
\bee\label{sigma}
\sigma(z)=\begin{cases} 0.1\sin(\pi z/2000)+0.1, \quad  0\leq z <1000, \\
           0.2, \qquad\qquad \qquad\qquad \qquad                      1000\leq z
          \end{cases}
\ene
which makes the potential $V(x)$ and gain-or-loss distribution $W(x)$ given by Eq.~(\ref{hg}), and solutions (\ref{solu2}) change, but it does not change the TOD coefficient
given by Eq.~(\ref{gau}). We consider the wave evolution of the solution $\psi(x,z)$ satisfied by Eq.~(\ref{nls2}) with Eqs.~(\ref{gau}), (\ref{hg}), and (\ref{sigma}) subject to the initial condition given by Eq.~(\ref{solu2}) such that we find the stable nonlinear modes by using the excitation, that is, we can smoothly excite one stable mode (see Fig.~\ref{stability-gau}(c)) to another stable mode (see Fig.~\ref{excite-gau}(b)). Moreover, if we  simultaneously consider the effect of varying $\beta_0(z)$ and $\sigma(z)$ given by  Eqs.~(\ref{beta0}) and (\ref{sigma}) then we also find the similar result (see Fig.~\ref{excite-gau}(c)).

\begin{figure}[!t]
	\begin{center}
		\vspace{0.05in}{\scalebox{0.75}[0.85]{\includegraphics{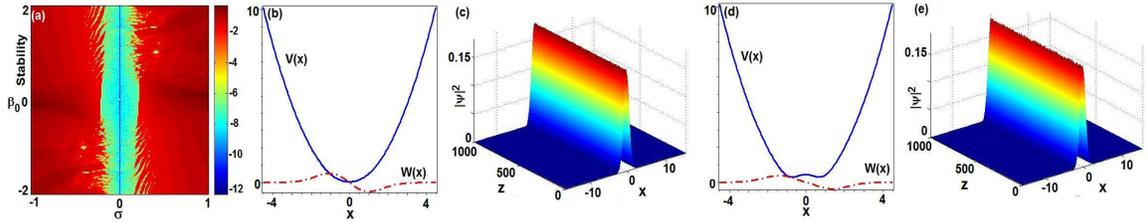}}}
				\end{center}
	\vspace{-0.15in}
	\caption{ {\bf Stability and dynamical behaviors with $g=1$.} (a) stable and unstable regions, (b) $\PT$-symmetric potential with sing-well $\sigma=\beta_0=0.1$,
 (c) stable nonlinear mode for $\sigma=\beta_0=0.1$ ( linear unbroken $\PT$-symmetric phase), (d) $\PT$-symmetric potential with double-well $\sigma=0.1,\, \beta_0=1$, (e) stable nonlinear mode for $\sigma=0.1,\, \beta_0=1$ (linear broken $\PT$-symmetric phase). } \label{stability-gau}
\end{figure}

\begin{figure}[!t]
	\begin{center}
		\vspace{0.05in}{\scalebox{0.45}[0.45]{\includegraphics{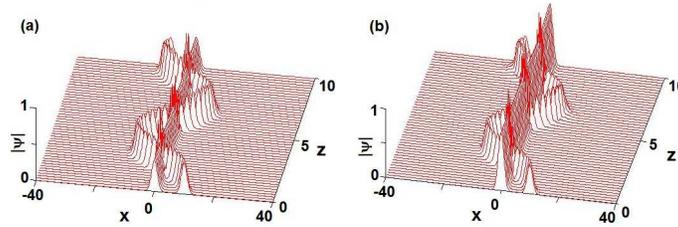}}}
				\end{center}
	\vspace{-0.15in}
	\caption{ {\bf  The interactions of bright solitons (\ref{solu2}) of Eq.~(\ref{nls2}).} (a) $\sigma=-0.1$, (b) $\sigma=0.2$. Other parameters are $g=1,\, \beta_0=0.1$.} \label{coll-gau}
\end{figure}

\begin{figure}[!t]
	\begin{center}
		\vspace{0.05in}{\scalebox{0.6}[0.6]{\includegraphics{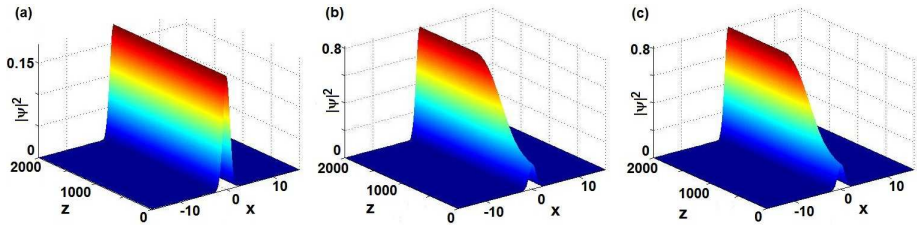}}}
				\end{center}
	\vspace{-0.15in}
	\caption{ {\bf Exciting stable nonlinear localized modes.} (a) $\sigma=0.1$ and $\beta_0(z)$ given by Eq.~(\ref{beta0}),
     (b) $\beta_0=0.1$ and $\sigma(z)$ given by Eq.~(\ref{sigma}), (c) $\beta_0(z)$ and $\sigma(z)$ given by  Eqs.~(\ref{beta0}) and (\ref{sigma}). } \label{excite-gau}
\end{figure}

\vspace{0.1in}
\noindent{\large\bf Discussion.} In conclusions, we have introduced some non-Hermitian (e.g., complex $\PT$-symmetric) potentials in the nonlinear Schr\"odinger equation with third-order dispersion. For the chosen physically interesting $\PT$-symmetric Scarff-II-like and harmonic-Gaussian potentials, we found exact analytical bright solitons of this
equation. In the presence of these $\PT$-symmetric potentials, we study the broken and unbroken of $\PT$-symmetric phases of the corresponding linear problem (third-order linear operator with complex potentials) for TOD and potential parameters such that we find the TOD parameter has a strong effect on the spectra (it only admit a few discrete spectra). We have studied the linear stability of exact bright solitons. In particular, we find the stable nonlinear modes for some control parameters for which even if the corresponding linear $\PT$-symmetric phase is broken. Moreover, we also investigate the problems of nonlinear modes excitations, which can excite initial nonlinear modes to reach stable modes. The method in this paper can also be extended to explore other higher-order or/and higher-dimensional NLS equations in the presence of non-Hermitian potentials and may open a new window to investigate similar problems. Our results may be useful to provide theoretical researchers and experimental scientists
with more new data about the $\PT$-symmetric nonlinear modes in higher-order nonlinear wave models.

\vspace{0.1in}
\noindent{\large\bf Methods}

\noindent{\bf Linear spectrum problem.} For Eq.~(\ref{nls}) in the absence of nonlinear term ($g=0$), we assume that $\psi(x,z)=\Phi(x)e^{-i\lambda z}$, then we have Eq.~(\ref{ls}), which with the $\PT$-symmetric potential (\ref{gps}), as $|x|\to \infty$, reduces to $-\left(\frac12\frac{d^2}{dx^2}+i\frac{\beta}{6} \frac{d^3}{dx^3}\right)\Phi(x)=\lambda\Phi(x)$,
whose characteristic equation is $\frac12\Lambda^2+\frac{i\beta}6 \Lambda^3+\lambda=0$, whose roots, in general, are complex numbers and complicated. For example, if $\lambda=1/(3\beta^2)$ and $\beta>0$ (without loss of generality), we have its three roots $\Lambda_1=i/\beta,\Lambda_{2,3}=(1\pm i\sqrt{3})/\beta$, which probably lead to the result that the corresponding eigenfunctions should satisfy periodic boundary conditions. Additionally, if $\beta$ depends on the space $x$, e.g., $\beta(x)=\beta_0 \exp(-x^2)$, and $V(x),\ W(x)$ are given by Eq.~(\ref{hg}), then we have $\beta(x),\, W(x)\to 0$ and $V(x)\to x^2/2$  as $|x|\to \infty$. Thus for this case Eq.~(\ref{ls}) reduces to
$\left(-\frac12\frac{d^2}{dx^2}+\frac{1}{2}x^2\right)\Phi(x)=0$ as $|x|\to \infty$, where the condition $\lambda\ll x^2$ is used, and we have the asymptotic solutions $\Phi(x)\sim e^{\pm x^2/2}$. Based on the standard conditions of wave function, we only take $\Phi(x)\sim e^{-x^2/2}$, which generally corresponds to zero boundary conditions and discrete spectra. Therefore, in order to verify these results, we use the Fourier collocation method~\cite{fm1, fm2, yang} to numerically study the above-mentioned linear spectrum problems and obtain the agreeable conclusions as ones by the theoretical analysis.

\vspace{0.05in}\noindent{\bf Nonlinear stationary modes.}  We consider the stationary solutions of Eq.~(\ref{nls}) in the form $\psi(x,z)=\phi(x)e^{-i\mu z}$, where
$\phi(x)$ is a complex field function  and $\mu$ the corresponding propagation constant. We have $\mu \phi+\frac{1}{2}\phi_{xx}+i\frac{\beta}{6}\phi_{xxx}-[V(x)+iW(x)]\phi+g|\phi|^2\phi=0.$ To study nonlinear modes of this equation we assume that  $\phi(x)=\rho(x)\exp[i\varphi(x)]$ with $\rho(x)$ and $\varphi(x)$ being real functions and separate the real and imaginary parts to yield
\bee
(\beta\varphi_x-3)\rho_{xx}+\beta\varphi_x\rho_x+[\beta\varphi_{xxx}+3\varphi_x^2-\beta\varphi_x^3-6V(x)+6\mu]\rho+6g\rho^3=0, \\
\beta\rho_{xxx}+(6\varphi_x-3\beta\varphi_x^2)\rho_x+3[\varphi_{xx}-\beta\varphi_{x}\varphi_{xx}-2W(x)]\rho=0.
\ene
For the given $\PT$-symmetric potential (\ref{gps}) we can find the exact bright solitons (\ref{solu}) of Eq.~(\ref{nls}). Similarly,
we can also find the solutions of Eq.~(\ref{nls2}) in the $\PT$-symmetric potential (\ref{hg}).

\vspace{0.1in}
\noindent{\bf Linear stability of nonlinear stationary modes.}  To further study the linear stability of the above-obtained nonlinear stationary solutions $\psi(x,z)=\phi(x)e^{-i\mu z}$ of  Eq.~(\ref{nls}), we considered a perturbed solution~\cite{stable, yang} $\psi(x,z)=\left\{\phi(x)+\epsilon \left[F(x)e^{i\delta z}\!+\!G^*(x)e^{-i\delta^* z}\right]\right\}e^{-i\mu z}$, where $\epsilon\ll 1$, $\delta$ and $F(x)$ and $G(x)$ are the eigenvalue and eigenfunctions of the linearized eigenvalue problem. We substitute the expression into Eq.~(\ref{nls}) and linearize it with respect to $\epsilon$ to yield the following linear eigenvalue problem
\begin{eqnarray} \label{st}
\left(\begin{array}{cc}   \tilde{L} & g\phi^2(x) \vspace{0.05in}\\   -g\phi^{*2}(x) & -\tilde{L}^* \\  \end{array}\right)
\left(  \begin{array}{c}    F(x) \vspace{0.05in} \\    G(x) \\  \end{array} \right)
=\delta \left(  \begin{array}{c}   F(x) \vspace{0.05in}\\    G(x) \\  \end{array}\right),
\label{stable}
\end{eqnarray}
where $\tilde{L}=\frac12\partial^2_x+i\frac{\beta}{6}\partial^3_x-[V(x)+iW(x)]+2g|\phi(x)|^2+\mu$. It is easy to see that
the $\PT$-symmetric nonlinear modes are linearly stable if $\delta$ has no imaginary component, otherwise they are linearly
unstable.

\vspace{0.1in}

\noindent{\large\bf Acknowledgments}\\
{\small This work was supported by the NSFC under Grant No.11571346 and the Youth Innovation Promotion Association CAS.}

\vspace{0.15in}

\noindent{\large\bf Author contributions} \\
Z.Y. conceived the idea and presented the overall theoretical analysis. Y.C. and Z.Y. discussed and performed
the numerical experiments. Z.Y. wrote the manuscript.

\vspace{0.15in}

\noindent{\large\bf Additional information}\\
Competing financial interests: The authors declare no competing financial interests.


\begin{thebibliography}{99} {\small

\begin{spacing}{0.5}

\bibitem{nls} Kivshar, Y. S. \& Agrawal, G. P. {\em Optical solitons: from fibers to photonic crystals}
(Academic Press, 2003).
\bibitem{ss3}  Agrawal, G. P. {\em Nonlinear fiber optics}, 4th edn ( Academic Press, 2006).

\bibitem{bec} Pitaevskii, L. \& Stringari, S. {\em Bose-Einstein condensation}, vol. 116 (Oxford University Press, 2003).

\bibitem{nls2} Kharif, C., Pelinovsky, E. \& Slunyaev, A. {\em Rogue waves in the ocean}
(Springer, 2009).

\bibitem{nls3} Kartashov, Y., Malomed, B. A. \& Torner, L.  Solitons in nonlinear lattices. {\it Rev. Mod. Phys.} {\bf 83}, 247 (2011).

\bibitem{yanctp10} Yan, Z. Financial rogue waves. {\it Commun. Theor. Phys.} {\bf 54}, 947-949 (2010).

 \bibitem{yanpla11} Yan, Z. Vector financial rogue waves. {\it Phys. Lett. A}  {\bf 375}, 4274-4279 (2011).

\bibitem{nls3a} Mihalache, D. Multidimensional localized structures in optics and
Bose-Einstein condensates: A selection of recent studies.
{\it Rom. J. Phys.} {\bf 59}, 295-312 (2014).

\bibitem{nls3b} Bagnato, V. S. {\it et al.} Bose-Einstein condensation: Twenty years after. {\it Rom. Rep. Phys.} {\bf 67}, 5-50 (2015).

\bibitem{s1} Malomed, B. A., Mihalache, D., Wise, F. \& Torner, L. Spatiotemporal optical solitons. {\it J. Opt. B: Quantum
Semiclass. Opt.} {\bf 7}, R53 (2005).

\bibitem{s2} Dudley, J. M. {\it et al.} Self-similarity in ultrafast nonlinear optics. {\it  Nature Phys.} {\bf 3}, 597-603 (2007).

\bibitem{s3} Serkin, V. N. \& Hasegawa, A. Novel soliton solutions of the nonlinear Schrodinger equation model.
  {\it Phys. Rev. Lett.} {\bf 85}, 4502 (2000).

\bibitem{s3a} Liang, Z. X., Zhang, Z. D. \& Liu, W. M. Dynamics of a bright soliton in Bose-Einstein condensates with time-dependent atomic
scattering length in an expulsive parabolic potential. {\it Phys. Rev. Lett.}  {\bf 94}, 050402 (2005).

\bibitem{s4} Belmonte-Beitia, J. {\it et al.} Localized nonlinear waves in systems with time- and space-modulated nonlinearities. {\it Phys. Rev. Lett.} {\bf 100}, 164102 (2008).

\bibitem{s5} Yan, Z. \& Konotop, V. V. Exact solutions to three-dimensional generalized nonlinear Schr\"odinger equations with varying potential and nonlinearities. {\it  Phys. Rev. E} {\bf  80}, 036607 (2009).

\bibitem{s6} Yan, Z., Konotop, V. V. \& Akhmediev, N. Three-dimensional rogue waves in nonstationary parabolic
potentials {\it Phys. Rev. E} {\bf 82}, 033610 (2010).

\bibitem{s7} Chen, Z. {\it et al.} Storage and retrieval of $(3+1)$-dimensional weak-light bullets and vortices in a coherent atomic gas. {\it Sci. Rep.} {\bf 5}, 8211 (2015).

\bibitem{bender1} Bender, C. M. \& Boettcher, S. Real spectra in non-Hermitian Hamiltonians having ${\cal PT}$ symmetry.
{\em Phys. Rev. Lett.} {\bf 80}, 5243-5246 (1998).

\bibitem{bender2} Bender, C. M. Making sense of non-Hermitian Hamiltonians. {\it Rep. Prog. Phys.}
{\bf 70}, 947-1018 (2007).

\bibitem{ziad} Musslimani, Z. H. {\it et al.}
Optical solitons in $\PT$ periodic potentials. {\it Phys. Rev. Lett.} {\bf 100}, 030402 (2008).


\bibitem{pt0} Chong, Y. D., Ge, L. \& Douglas Stone, A. D. $\PT$-symmetry breaking and laser-absorber modes in optical scattering systems. {\it Phys. Rev. Lett.} {\bf 106}, 093902 (2011).

\bibitem{pt1} Abdullaev, F. Kh. {\it et al.} Solitons in $\PT$-symmetric nonlinear lattices. {\it Phys. Rev. A}
 {\bf 83}, 041805R (2011).

\bibitem{pt2} Li, K. \& Kevrekidis, P. G. $\PT$-symmetric oligomers: Analytical solutions, linear stability, and nonlinear dynamics. {\it Phys. Rev. E}  {\bf 83}, 066608 (2011).

\bibitem{pt3} Nixon, S., Ge, L. \& Yang, J. Stability analysis for solitons in $\PT$-symmetric optical lattices.
{\it Phys. Rev. A} {\bf 85}, 023822 (2012).

\bibitem{pt4} Achilleos, V. {\it et al.} Dark solitons and vortices in $\PT$-symmetric nonlinear media: From spontaneous symmetry breaking to nonlinear $\PT$ phase transitions.
 {\it Phys. Rev. A} {\bf 86}, 013808 (2012).

\bibitem{pt5}  Zezyulin, D. A. \& Konotop, V. V. Nonlinear modes in the harmonic $\PT$-symmetric potential. {\it Phys. Rev. A} {\bf 85}, 043840 (2012).

\bibitem{pt6}  Cartarius, H. \& Wunner, G. Model of a $\PT$-symmetric Bose-Einstein condensate in a $\delta$-function double-well potential. {\it Phys. Rev. A} {\bf 86}, 013612 (2012).

\bibitem{pt7}  D. A. Zezyulin, D. A. \& Konotop, V. V. Nonlinear modes in finite-dimensional $\PT$-symmetric systems. {\it Phys. Rev. Lett.}  {\bf 108}, 213906 (2012).

\bibitem{pt72} Luo, X. {\it et al.} Pseudo-parity-time symmetry in optical systems. {\it Phys. Rev. Lett.} {\bf 110}, 243902 (2013).

\bibitem{pt73}  Hang, C., Huang, G. \& Konotop, V. V. $\PT$ symmetry with a system of three-level atoms. {\it Phys. Rev. Lett.}
 {\bf 110}, 083604 (2013).

\bibitem{pt8} Lumer, Y. {\it et al.} Nonlinearly induced $\PT$ transition in photonic systems. {\it Phys. Rev. Lett.} {\bf 111}, 263901 (2013).

\bibitem{pt82} Pickton, J. \& Susanto, H. Integrability of $\PT$-symmetric dimers. {\it Phys. Rev. A} {\bf 88}, 063840 (2013).

\bibitem{pt81} Yan, Z. Complex $\PT$-symmetric nonlinear Schr\"odinger equation and Burgers equation. {\it Phil. Trans. R. Soc. A} {\bf 371}, 20120059 (2013).

\bibitem{pt83} Saleh, M. F., Marini, A. \& Biancalana, F. Shock-induced $\PT$-symmetric potentials in gas-filled photonic-crystal fibers. {\it Phys. Rev. A} {\bf 89}, 023801 (2014).

\bibitem{pt84} Yang, J. Partially PT symmetric optical potentials with all-real spectra and soliton families in multidimensions. {\it Opt. Lett.}
{\bf 39}, 1133 (2014).

\bibitem{pt9} Yan, Z., Wen, Z. \& Konotop, V. V.  Solitons in a nonlinear Schr\"odinger equation with $\PT$-symmetric potentials
and inhomogeneous nonlinearity: stability and excitation of nonlinear modes. {\it Phys. Rev. A} {\bf 92}, 023821 (2015).

\bibitem{pt11} Makris, K. G. {\it et al.} Constant-intensity waves and their modulation instability in non-Hermitian potentials. {\it Nat. Commun.} {\bf 6}, 7257 (2015).

\bibitem{pt10} Yan, Z., Wen, Z. \& Hang, C. Spatial solitons and stability in self-focusing and defocusing Kerr nonlinear media with generalized parity-time-symmetric Scarff-II potentials. {\it Phys. Rev. E} {\bf 92}, 022913 (2015).

\bibitem{pt10a}  Wen, Z. \& Yan, Z. Dynamical behaviors of optical solitons in parity-time (PT) symmetric sextic anharmonic double-well potentials. {\it Phys. Lett. A} {\bf 379}, 2025-2029 (2015).

\bibitem{pt12} Kartashov, Y. V., Konotop, V. V. \& Torner, L. Topological states in partially-$\PT$-symmetric azimuthal potentials. {\it Phys. Rev. Lett.} {\bf 115}, 193902 (2015).

\bibitem{pt13} Xu, H. {\it et al.} Nonlinear $\PT$-symmetric models bearing exact solutions. {\it Rom. J. Phys.} {\bf 59}, 185-194 (2014).

\bibitem{pt14} Liu, B., Li, L., \& Mihalache, D. Vector soliton solutions in $\PT$-symmetric coupled waveguides and their relevant properties. {\it Rom. Rep. Phys.} {\bf 67}, 802-818 (2015).

\bibitem{exp1} Guo, A. {\it et al.} Observation of $\PT$-symmetry breaking in complex optical potentials.
{\it Phys. Rev. Lett.} {\bf 103}, 093902 (2009).

\bibitem{exp2} R\"uter, C. E. {\it et al.} Observation of parity-time symmetry in optics. {\it Nature Phys.} {\bf 6,} 192-195 (2010).

\bibitem{exp3} Regensburger, A. {\it et al.} Parity-time synthetic photonic lattices. {\it Nature} {\bf 488}, 167-171 (2012).

\bibitem{exp4} Peng, B. {\it et al.} Parity-time-symmetric whispering gallery microcavities. {\it Nature Phys.} {\bf 10}, 394-398 (2014).


\bibitem{exp6} Hodaei, H. {\it et al.} Parity-time-symmetric microring lasers. {\it Science}  {\bf 346}, 975 (2014).

\bibitem{exp7} Wimmer, M. {\it et al.} Observation of optical solitons in $\PT$-symmetric lattices. {\it Nature  Commun.}  {\bf 6}, 7782 (2015).


\bibitem{ss} Kodama, Y. Optical solitons in a monomode fiber. {\it J. Stat. Phys.} {\bf 39}, 597 (1985).

\bibitem{ss2} Kodama, Y. \& Hasegawa, A. Nonlinear pulse propagation in a monomode dielectric guide. {\it IEEE J. Quantum Electron.} {\bf 23}, 510 (1987).

\bibitem{yanjop} Yan, Z \& C. Dai. Optical rogue waves in the generalized inhomogeneous higher-order nonlinear Schr\"odinger equation with modulating coefficients.
  {\it J. Opt.}, {\bf 15}, 064012 (2013).


\bibitem{3nls3} Wang, S. F. {\it et al.} Optical event horizons from the collision of a soliton and its own dispersive wave.
{\it Phys. Rev. A} {\bf 92}, 023837 (2015).

\bibitem{3nls1} Bhat, N. A. R. \& Sipe, J. E. Optical pulse propagation in nonlinear photonic crystals. {\it Phys. Rev. E} {\bf 64}, 056604 (2001).

\bibitem{3nls2} Colman, P. {\it et al.} Temporal solitons and pulse compression in photonic crystal waveguides. {\it Nature Photon} {\bf 4}, 862-868 (2010).


\bibitem{ss4} Mihalache, D. {\it et al.} Painlev\'e analysis and bright solitary waves of the higher-order nonlinear Schr\"odinger equation containing third-order dispersion and self-steepening term, {\it Phys. Rev. E} {\bf 56}, 1064 (1997).


\bibitem{raman} Robertson, S. \& Leonhardt, U. Frequency shifting at fiberoptical
event horizons: The effect of the Raman deceleration, {\it Phys. Rev. A} {\bf 81}, 063835 (2010).

\bibitem{rp} P\"oschl, G. \& Teller, E. Bemerkungen zur quantenmechanik des anharmonischen oszillators, {\it Z. Phys}. {\bf 83}, 143-151 (1933).

\bibitem{sc} Ahmed, A. Real and complex discrete eigenvalues in an exactly solvable one-dimensional complex $\PT$-invariant potential, {\it Phys. Lett. A}, {\bf 282}, 343-348 (2000).

\bibitem{fm1} Trefethen, L. N. {\it Spectral methods in Matlab} (SIAM, 2000).

\bibitem{fm2} Shen, J. \& Tang, T. {\it Spectral and high-order methods with applications}, vol. 3 (Science Press, 2006).

\bibitem{yang} Yang, J. {\it Nonlinear waves in integrable and nonintegrable systems} (SIAM, 2010).

\bibitem{stable} Kuznetsov, E. A. {\it et al.} Soliton stability in plasmas and hydrodynamics, {\it Phys. Rep}. {\bf 142}, 103-165 (1986).

\end{spacing}
}
\end{thebibliography}
\end{document}